\documentclass[prr,letterpaper,twocolumn,aps,floatfix]{revtex4-2}
\usepackage{tabularx}
\usepackage{array}
\usepackage{graphicx}
\usepackage{xcolor}
\usepackage{amsmath}
\usepackage{amsfonts}
\usepackage[small,bf]{subfigure}
\newcommand{\beq}{\begin{equation}}
\newcommand{\eeq}{\end{equation}}

\newcommand{\br}{{{\bf{r}}}}
\newcommand{\bR}{{{\bf{R}}}}

\newcommand{\bb}{{\bf{b}}}

\newcommand{\beqa}{\begin{eqnarray}}
\newcommand{\eeqa}{\end{eqnarray}}

\newcommand{\ra}{\rightarrow}

\begin{document}
\title{Anderson localization in topological metals}
\author{A.A. Burkov}
\affiliation{Department of Physics and Astronomy, University of Waterloo, Waterloo, Ontario 
N2L 3G1, Canada} 
\affiliation{Perimeter Institute for Theoretical Physics, Waterloo, Ontario N2L 2Y5, Canada}
\date{\today}
\begin{abstract}
We address the question of whether topological metals (TM) retain their nontrivial characteristics in the presence of disorder. 
While in the clean case, the issue is what protects the gaplessness of the spectrum, i.e. the band-touching nodes in TM, 
this question needs to be reformulated in the presence of disorder. The concept of gaplessness loses its sharp meaning in this case, since 
spectral gaps can be filled in by disorder-induced states. What is still meaningful is entanglement: one may ask whether the long-range 
entanglement of the clean gapless TM survives in the presence of disorder or, in other words, whether nontrivial topology protects the TM from Anderson 
localization. We demonstrate that such a protection exists in the case of three-dimensional (3D) Weyl semimetals, but not in 2D and 3D (type-I) Dirac or 
3D nodal line semimetals.  We establish this by combining the idea of unquantized anomaly as the topological response of TM with the 
decorated domain wall construction, used previously to discuss average symmetry-protected topological insulating phases. 

\end{abstract}
\maketitle
\section{Introduction}
\label{sec:1}
The interplay of disorder and electronic structure topology is an important theme in the modern theory of quantum matter. 
One of the characteristic features of topological insulators (TI)~\cite{Hasan10,Qi11}, for example, is the fact that their edge states are often immune to Anderson 
localization, as long as disorder is not strong enough to eliminate the distinction between the topological and ordinary insulator. 
This property is closely related to the fact that the surface states of TI are anomalous, in the sense that they may not be realized as bulk states in the same dimension. 

The situation in bulk topological metals (TM), such as Weyl, Dirac and nodal line semimetals~\cite{Volovik03,Volovik07,Murakami07,Wan11,Burkov11-1,Burkov11-2,Weyl_RMP,Ozawa26}, 
is more complex. 
In clean bulk TM, while there is no quantized topological invariant, which requires gapless states on the surface of TI, gaplessness is still protected by an ``unquantized anomaly", 
as discussed in Refs.~\cite{Gioia21,Wang21,Wang24,Hughes24}. 
This is a topological term, but with a continuously tunable coefficient (related to, e.g. the distance between the band-touching nodes in momentum space), which is not gauge invariant 
on its own. Such a term, while itself an equilibrium contribution, arising from all filled states and thus stable to small perturbations, requires gapless modes to be present at the Fermi energy 
to restore gauge invariance. This is a generalization of the Lieb-Schultz-Mattis constraint and Luttinger's theorem in ordinary metals~\cite{Luttinger60,LSM,Oshikawa00,Hastings04}, where a fractional electron filling per unit cell requires a 
Fermi surface, whose volume is directly determined by the fractional part of the filling. In TM the Luttinger volume is zero, but the unquantized anomaly, expressing e.g. fractional 
Hall conductance per atomic plane in a magnetic Weyl semimetal, plays an analogous role.

However, the presence of disorder complicates matters. Unquantized anomalies in clean TM always involve crystal symmetry (such as translation, rotation, mirror, etc.) gauge fields, 
which expresses the fact that these symmetries are essential in protecting the gaplessness of the spectrum in clean TM. Disorder violates crystal symmetries in any given sample, thus 
seemingly eliminating the protection. However, as long as a given symmetry is restored upon averaging over disorder realizations or, equivalently, at long length scales, unquantized anomaly terms and the corresponding 
observable responses are still meaningful. This is analogous to the idea that symmetry-protected topological (SPT) phases still exist in the presence of disorder in the average 
sense, even when disorder violates the protecting symmetry locally in a given realization~\cite{Ma22}. 

Another issue, however, is that the concept of gaplessness loses its sharp meaning in disordered systems, since all spectral gaps can be filled in by disorder-induced states. 
Thus asking whether unquantized anomaly requires a gapless spectrum is, in general, a meaningless question in the presence of disorder. 
What remains meaningful is entanglement: one may ask whether the unquantized anomaly requires long-range entanglement, which in the clean case is present in any gapless 
(or topologically ordered) system of fermions. 

We know that the answer in ordinary metals is generally no: a sufficient amount of disorder always (with some exceptions, like the two-dimensional electron gas at the integer quantum Hall plateau transition, which actually shares a lot of similarities with TM~\cite{Yi25}) 
leads to an Anderson insulator even at a fractional filling of electrons 
per unit cell, which is a short-range entangled state due to the finite localization length.  
The question we ask in this paper is whether the same is true in TM. 

The answer turns out to depend on the type of the corresponding topological response, or unquantized anomaly. 
In particular, we demonstrate that the topology of three-dimensional (3D) Weyl semimetals requires long-range entanglement and thus prohibits Anderson localization~\cite{Yi24,Yi25,Burkov25}. 
In contrast, 3D type-I and 2D Dirac as well as 3D nodal-line semimetals are not 
protected against localization as their unquantized anomalies are consistent with a short-range entangled state in the presence of disorder. 

We demonstrate this using a generalization of the decorated domain wall construction, which was used in Ref.~\cite{Ma22} to discuss average SPT. 
It was shown in that case that the average SPT exists as long as the domain wall spatial dimension is greater than zero (i.e. they are not points). 
We will demonstrate that the condition for a nontrivial TM in the presence of disorder is more restrictive and requires that the topological domain wall bound states 
are themselves protected from localization. This condition is satisfied only by 3D Weyl semimetals. 
In contrast, the topological response of 2D, type-I 3D Dirac and 3D nodal line semimetals is in fact compatible with a short-range entangled Anderson insulator in the presence of disorder
and the decorated domain wall approach gives an explicit construction of such localized insulating states. 

The rest of the paper is organized as follows. 
In Section~\ref{sec:2}, to set the stage for the decorated domain wall construction, we review the theory of unquantized anomalies in TM. 
In Section~\ref{sec:3} we apply this theory to the case of disordered TM using the decorated domain wall construction. 
We conclude in Section~\ref{sec:4} with an overview of our results. 
\section{Unquantized anomalies in TM}  
\label{sec:2}
Since our arguments depend crucially on the concept of the unquantized anomaly and the corresponding topological terms, and to keep the paper self-contained, we will start with a brief review  of the results of Refs.~\cite{Gioia21,Wang21,Hughes24}, where this concept was introduced and developed. 
\subsection{Unquantized chiral anomaly}
\label{sec:2.1}
As demonstrated in Ref.~\cite{Wang21}, the type of anomaly that defines the topological response in a given TM depends on the codimension of the low-energy mode manifold in momentum space. The anomaly is chiral when the codimension is odd, and it is parity when the codimension is even. 
For a 3D point-node semimetal the codimension is 3 and thus it is the chiral anomaly that is operative. 

Let us start from the simplest and most important example of a magnetic Weyl semimetal with a pair of band-touching nodes, introduced theoretically in Refs.~\cite{Burkov11-1,Burkov11-2} 
and experimentally realized in \cite{Belopolski25}. 
If the Weyl nodes are separated by momentum $2 Q$ along the $z$-axis, then the topological response is the Hall conductivity
\beq
\label{eq:1}
\sigma_{xy} = \frac{2 Q}{2 \pi} \frac{e^2}{h} = \frac{1}{a} \frac{\lambda}{2 \pi}, 
\eeq
where we will switch to $\hbar = c = e = 1$ units henceforth, $a$ is the lattice constant in the $z$-direction
and $\lambda \in [0,1]$ is a dimensionless measure of the separation between the Weyl nodes relative to 
the total size of the Brillouin zone (BZ)
\beq
\label{eq:2}
\lambda = \frac{2 Q}{2 \pi/a}. 
\eeq
 Eq.~\eqref{eq:1} may then be interpreted as a statement that the Weyl semimetal is characterized by a noninteger Hall conductance $g_{xy} = \lambda/2 \pi$ per atomic plane, normal 
 to the primitive reciprocal lattice vector $\bb^z = \frac{2 \pi}{a} \hat z$.
 
 While the Hall conductivity of Eq.~\eqref{eq:1} is not dimensionless (in units of $e^2/h$) and depends on a nonuniversal and variable lattice scale $a$, the conductance per atomic plane $g_{xy}$ is universal and may, at least in principle, be precisely measured and fixed. 
 This suggests that it is $g_{xy}$, or the corresponding number $\lambda$, that should really be used as the primary characteristic of the Weyl semimetal state. 
 This idea may be formalized using the concept of a translation gauge field, which may be thought of as a mathematical device designed to reinterpret the topological linear response of the Weyl semimetal in terms of $g_{xy}$ rather than $\sigma_{xy}$~\cite{Volovik19,Song21,Nissinen21}. 
 This concept is based on the dual description of a crystal in terms of intersecting families of crystal planes. 
 
 A perfect 3D crystal may be characterized by three independent families of crystal planes, which may be viewed as solutions of the equations
 \beq
 \label{eq:3}
 \theta^i(\br, t) = \bb^i \cdot \br = \bb^i \cdot \bR = 2 \pi n^i. 
 \eeq
 Here the index $i = 1, 2, 3$ labels the three families of planes (for simplicity, we will implicitly assume cubic symmetry with $1, 2, 3 = x, y, z$, but all the physics is of course independent 
 of this assumption), $\bb^i$ are the basis vectors of the reciprocal space, $\bR$ are Bravais lattice vectors and $n^i$ are integers. 
 Eq.~\eqref{eq:3} says that the phase $\theta^i$ winds by $2 \pi$ between every nearest-neighbor pair of planes in a family. The total winding in a given direction counts the total number 
 of planes (unit cells) in this direction. 
 In a perfect crystal, the gradients of the phase functions $\theta^i$ give the corresponding primitive reciprocal lattice vectors
 \beq
 \label{eq:4}
 b^i_j = \partial_j \theta^i. 
 \eeq
 This may be generalized to an arbitrarily distorted crystal, including a crystal with dislocations and time-dependent distortions, by introducing a one-form 
 \beq
 \label{eq:5}
 e^i = \frac{1}{2 \pi} d \theta^i = \frac{1}{2 \pi} \partial_{\mu} \theta^i dx^{\mu}, 
 \eeq
 where $\mu$ may include time as well as spatial coordinates. 
 In a crystal without dislocations this form is closed
 \beq
 \label{eq:6}
 d e^i = \frac{1}{2} (\partial_{\mu} e^i_{\nu} - \partial_{\nu} e^i_{\mu}) d x^{\mu} d x^{\nu} = 0. 
 \eeq
On the other hand, if a dislocation with a Burgers vector along $\bb^i$ is present, the integral of the one-form $e^i$ along a cycle $C$, enclosing the dislocation line, is
\beq
\label{eq:7}
\int_C e^i = \pm 1. 
\eeq
If $C_i$ is the $i$th fundamental noncontractible cycle, going around the whole sample (with periodic boundary conditions) in the direction $\bb^i$, then 
\beq
\label{eq:8} 
\int_{C_i} e^i = N_i, 
\eeq
where $N_i$ is the total number of unit cells (total winding of the phase function $\theta^i$) in this direction. 
 If we focus on the topological information only, encoded in Eqs.~\eqref{eq:7} and \eqref{eq:8}, and discard the geometric information, contained in the space and time-dependent local 
 value of the components $e^i_{\mu}$, we may view the one-form $e^i$ as a translation symmetry gauge field, defined by these equations. 
 A useful way to think about this is based on the fact that the phase functions $\theta^i$ are defined modulo $2 \pi$. This implies that the integers $n^i$ in Eq.~\eqref{eq:3} may be 
 relabelled arbitrarily (a gauge transformation), but this relabelling leaves Eqs.~\eqref{eq:7} and \eqref{eq:8} invariant. 

The concept of a translation gauge field allows us to express the fractional Hall conductance per atomic plane of a Weyl semimetal as a topological term
\beq
\label{eq:9}
S = i \frac{\lambda}{4 \pi} \int e^z \wedge A \wedge d A. 
\eeq
This looks like a standard topological term, except that the coefficient is not properly quantized, which makes Eq.~\eqref{eq:9} not gauge invariant under large 
gauge transformations of $A$. This means that this term must be accompanied by gapless modes (Weyl nodes), which restore the gauge invariance, see Ref.~\cite{Gioia21}
for more details on this. 

Another reason this viewpoint is useful is that it allows one to make a simple and transparent connection to 't Hooft anomalies. This is useful on its own and for deriving such responses
in cases when it is less obvious what they are. 
 To connect Eq.~\eqref{eq:9} with the standard chiral 't Hooft anomaly we note the following. 
If we focus only on the low-energy modes, a Weyl semimetal appears to have a larger symmetry group than it actually has. 
Its true microscopic symmetry group is $U(1) \times \mathbb{Z}$, where $U(1)$ corresponds to the electric charge conservation and 
$\mathbb{Z}$ to lattice translations (other discrete crystal symmetries, such as rotations, are irrelevant here). 
At low energies, however, the symmetry group appears to be $U(1)_L \times U(1)_R$, which corresponds to separate conservation of Weyl fermions
of left (L) and right (R) chirality. 
This symmetry group, if it were the true microscopic symmetry, would be anomalous, in the sense that it could not be realized in any 3D lattice model, but 
could only appear on the surface
of a four-dimensional (4D) topological insulator, described by the following topological field theory
\beq
\label{eq:10}
S = \frac{i}{6} \frac{1}{(2 \pi)^2} \int (A_R \wedge d A_R \wedge d A_R - A_L \wedge d A_L \wedge d A_L),
\eeq
where $A_{R,L}$ are $U(1)$ gauge fields, corresponding to the separate $U(1)_{R, L}$ symmetries. 
We now write $A_R = A + \tilde Q \tilde A$ and $A_L = A - \tilde Q \tilde A$, where $\tilde A$ is a chiral $U(1)$ gauge field, which 
couples antisymmetrically to fermions of opposite chirality and $\tilde Q$ is the corresponding ``charge". 
This gives
\beq
\label{eq:11}
S = i \tilde Q \frac{1}{(2 \pi)^2} \int \tilde A \wedge d A \wedge d A.
\eeq
On the 3D boundary of this 4D topological insulator, Eq.~\eqref{eq:11} becomes
\beq
\label{eq:12}
S = i \tilde Q \frac{1}{(2 \pi)^2} \int \tilde A \wedge A \wedge d A.
\eeq
If we now identify $\tilde Q = Q a = \pi \lambda$, i.e. the dimensionless magnitude of the momentum of the two Weyl points 
(i.e. translational symmetry charge) and 
$\tilde A = e^z$, we get precisely Eq.~\eqref{eq:9}, which describes the topological response of the physical 3D Weyl semimetal. 
The reason Eq.~\eqref{eq:9} can exist in a stand-alone 3D system, rather than a boundary of a 4D topological insulator, is that the physical translational 
symmetry group is $\mathbb{Z}$ rather than $U(1)$ and, moreover, translation is not a truly ``internal" symmetry, like chiral charge conservation. 

This connection to the 't Hooft anomaly of the emergent low-energy symmetry allows easy generalization to other types of TM. 
Consider now a TR-invariant Weyl semimetal with band-touching nodes, which we for simplicity take to be on the $z$-axis. 
The minimal number of such nodes is four in this case, i.e. two pairs of opposite-chirality nodes related by TR. 
Taking the nodes to be at momenta $k_z = \pm(Q \pm \delta Q)$, we may generalize the low energy symmetry 't Hooft anomaly argument in Eqs.~\eqref{eq:10}-\eqref{eq:12} 
to obtain the following topological term, describing the response of a TR-invariant Weyl semimetal
\beq
\label{eq:13}
S = i \frac{\lambda}{2} \int e^z \wedge d e^z \wedge A, 
\eeq
where $\lambda = 2 Q 2 \delta Q a^2/2 \pi^2$.
This describes a fractional electric charge of $\lambda$ per unit cell, induced on a screw dislocation with the Burgers vector along the $z$-direction. 
In terms of the Weyl node BZ geometry, $\lambda/2$ has the meaning of the Weyl node topological charge quadrupole moment, in units of the square of the primitive 
reciprocal lattice vector~\cite{Hughes24}. 

Using similar arguments, the topological response for a type-I Dirac semimetal with a pair of Dirac nodes on the $z$-axis, separated by momentum $2 Q$ and protected by an 
$n$-fold rotational symmetry with respect to the $z$-axis, may be derived. 
One obtains the response 
\beq
\label{eq:14}
S = i \frac{2 \pi q}{n} \frac{\lambda}{2 \pi} \int e^z \wedge c_n \wedge d A. 
\eeq
Here $c_n$ is a gauge field for the $n$-fold rotational symmetry, defined by 
\beq
\label{eq:15}
\oint c_n = m,
\eeq
for a cycle enclosing a disclination line with the Frank angle $\Omega_m = 2 \pi m/n$. 
$2 \pi q /n$, with an integer $q = 0, 1, \ldots, n-1$, is the angular momentum, carried by the mass term, that gaps out the Dirac nodes when the $n$-fold rotational symmetry is broken, 
and $\lambda = 2 Q a/2 \pi$ is the separation between the Dirac nodes in units of the primitive reciprocal lattice vector. 
The angular momentum of the mass term enters in Eq.~\eqref{eq:14} because this is the difference between the two rotation eigenvalues at the Dirac points, whose degeneracy 
is lifted by the mass term. This is analogous to the $2 Q a$ coefficient in the unquantized anomaly of the magnetic Weyl semimetal, which is again the linear momentum, carried by the
mass term, mixing the Weyl nodes. 
We refer the reader to Ref.~\cite{Gioia21} for further details of the derivation of Eq.~\eqref{eq:14} for a specific microscopic model of a type-I Dirac semimetal, protected by $C_4$ symmetry. 
\subsection{Unquantized parity anomaly}
\label{sec:2.2}
Let us now move on to the unquantized parity anomaly. This is operative for 3D line-node semimetals and 2D point-node semimetals. 
Let us start from the 2D point-node semimetal (sometimes called 2D Weyl semimetal, although this is an abuse of terminology)~\cite{Zyuzin11,Burkov18-2,Stemmer23,2DWeyl}. 
This is a nondegenerate point-node 2D semimetal with a pair of 2D Dirac nodes, separated in momentum space. 
The nodes are protected by translational symmetry as well as mirror symmetry. The simplest realization is a 3D TI thin film with an in-plane magnetization, 
normal to a mirror line~\cite{Zyuzin11,Burkov18-2,Stemmer23}. 

Following the same logic as in the case of the unquantized chiral anomaly, we note that at low energies, the 2D Dirac semimetal has separate conservation of the two 
species of Dirac fermions. This situation may not be realized in a stand-alone 2D system and may only appear on the surface of a 3D mirror-symmetric TI, described by the topological term
\beqa
\label{eq:16}
S&=&\frac{i \theta}{2 (2 \pi)^2} \int d(A + Q a e^x) \wedge d(A + Q a e^x) \nonumber \\
&-&\frac{i \theta}{2 (2 \pi)^2} \int d(A - Q a e^x) \wedge d(A - Q a e^x), 
\eeqa
where $k_x = \pm Q$ are the momenta of the two Dirac points and $\theta = \pm \pi$ is the topological theta-angle, pinned to this value by the mirror symmetry. 
The arbitrariness of the sign reflects the fact that $\theta$ changes sign under the mirror symmetry transformation and is defined modulo $2 \pi$. 
This leaves the following mixed topological response on the 2D boundary
\beq
\label{eq:17}
S = \pm i \frac{\lambda}{2} \int e^x \wedge d A, 
\eeq
where $\lambda = 2 Q a/2 \pi$. 
Following earlier logic, this is the topological response of a 2D Dirac semimetal with a pair of Dirac nodes at $k_x = \pm Q$, protected by mirror symmetry. 
Microscopically, this describes fractional electric polarization in the $y$-direction, i.e. perpendicular to a mirror line. 
A specific sign of the fractional polarization is selected by an infinitesimally small mirror symmetry breaking mass.

Why does such a response require a gapless state? Mirror symmetry reverses the polarization, while a short-range-entangled insulator has a well-defined polarization modulo the polarization quantum. A mirror-symmetric short-range-entangled state must therefore satisfy $P_y=-P_y$ modulo the polarization quantum. For the response Eq.~\eqref{eq:17}, 
this permits only $\lambda=0$ or $1$, whereas a continuously tunable $0<\lambda<1$ requires gapless modes, which in the present case form a pair of Dirac nodes.

This may be easily generalized to the case of the 3D nodal line semimetal, protected by mirror symmetry. In this case one obtains
\beq
\label{eq:18}
S = \pm i \frac{\lambda}{2} \int e^x \wedge e^y \wedge d A, 
\eeq
where $\lambda = V_F a^2/(2 \pi)^2$ and $V_F$ is the area in momentum space, enclosed by the nodal line, see Ref.~\cite{Wang21} for details of the derivation. 

\section{Decorated domain wall construction}
\label{sec:3}
Now we will use the unquantized anomaly theory of the topological response in TM to address the question of Anderson localization in the presence of disorder. 
We will use the following logic. 
Disorder locally violates the crystalline symmetries, protecting the band-touching nodes of the clean semimetals. This generates mass terms $m(\br)$, which may be scalar, complex 
or vector, depending on the symmetry. We assume, however, that the symmetry is restored on long length scales (or, equivalently, by averaging over many realizations). 
This implies that the masses $m(\br)$ have to average to zero on long length scales. This, in turn, implies proliferation of topological defects of the mass term spatial textures, which may carry 
gapless modes due to dimensional descent from the bulk unquantized anomaly terms~\cite{CallanHarvey}. 
The question of the possibility of Anderson localization of a given TM then reduces to whether the corresponding network of gapless modes may be localized or not. 
\subsection{Unquantized chiral anomaly}
\label{sec:3.2}
Let us start again from the magnetic Weyl semimetal case, whose unquantized anomaly response is given by Eq.~\eqref{eq:9}. 
Disorder will locally break translational symmetry, mixing the two Weyl nodes and gapping them out. The corresponding mass term is complex and 
carries momentum $2 Q$, i.e. the distance between the nodes in momentum space. 
This implies that, if we denote the phase of this random mass by $\theta(\br)$, the corresponding topological term may be obtained from Eq.~\eqref{eq:9} by the replacement
\beq
\label{eq:19}
2 Q a e^z = 2 \pi \lambda e^z \ra d \theta, 
\eeq
which gives
\beq
\label{eq:20}
S = \frac{i}{8 \pi^2} \int d\theta \wedge A \wedge d A. 
\eeq

Now let us demonstrate that this action implies 1D chiral modes, bound to vortex lines in $\theta$ (we will ignore here lattice commensuration effects, making $\theta$ discrete, 
these are not important for our argument). 
Indeed, let us make a gauge transformation $A \ra A + d \Lambda$. The corresponding change of the action is
\beq
\label{eq:21}
\delta S = \frac{i}{8 \pi^2} \int d \theta \wedge d \Lambda \wedge d A, 
\eeq
In the presence of a vortex line such that $\oint d \theta = 2 \pi$ on any cycle enclosing the line, this gives
\beq
\label{eq:22}
\delta S = \frac{i}{4 \pi} \int \Lambda d A, 
\eeq
where the integral is taken on the vortex worldsheet.  
The action thus fails to be gauge invariant in the presence of a vortex. 

It is easily seen that this failure of gauge invariance, or anomaly, is identical to the chiral anomaly at the edge of a 2D Chern insulator, described by the topological term
\beq
\label{eq:23}
S = \frac{i}{4 \pi} \int A \wedge d A. 
\eeq
In other words, each vortex line may be viewed as an edge of a 2D Chern insulator, which may be thought of as living on the 2D branch surface of the phase $\theta$, where it 
jumps by $2 \pi$. 
This anomaly is known to be cancelled by the 1D chiral edge mode of the 2D Chern insulator. 
It follows that vortex lines in the phase angle $\theta$ of the random mass term bind 1D chiral modes. 

To restore translational symmetry, such vortex lines must proliferate through the sample. 
Indeed, since the magnitude of the mass is always positive, we may simply take it to be constant (at least away from the vortex cores). 
Then, a translation by a lattice constant along $z$ acts on the symmetry-breaking mass as 
\beq
\label{eq:23.5}
m({\bf r}) \ra m({\bf r}) e^{2 i Q a}, 
\eeq
which means that statistical restoration of the translational symmetry is equivalent to a $U(1)$ symmetry (again ignoring commensuration effects) of the phase $\theta$. 
This means that the phase must not have long range order in a typical macroscopic sample. 
In 3D, such a disordered phase of a $U(1)$ field is a vortex loop proliferated phase, which in the thermodynamic limit means a system-spanning vortex network. 

Each vortex line carries a single chiral mode, whose propagation direction is fixed by the oriented vorticity of the line. A possible localized state could arise if the directed trajectories were all reconnected into finite closed loops, as in a localized phase of the Chalker-Coddington network~\cite{Chalker88,Chalker95}. However, this would contradict the long-length-scale restoration of translational symmetry, which excludes such a globally localized loop configuration, as explained above. 
One can make this statement more precise using the following argument, which avoids referring to specific properties of the network other than the chiral modes themselves. 
Suppose the network were localized with a finite localization length $\xi$, and coarse grain the system on length scales $L\gg\xi$. In an Anderson insulator no extended electronic states can remain as intrinsic long-distance degrees of freedom after such coarse graining. Statistical translational symmetry, however, is preserved under coarse graining. The coarse-grained phase 
$\theta$ must therefore remain in the symmetry-restored, vortex-proliferated phase, with vortex defects present on arbitrarily long length scales. The anomaly descent expressed by 
Eq.~\eqref{eq:22} is unchanged under coarse graining, so every elementary vortex of the coarse-grained phase still carries a single chiral mode. Thus extended chiral states necessarily survive on scales $L\gg\xi$, contradicting the assumption of a finite localization length.

This means that the unquantized topological response of a magnetic Weyl semimetal is inconsistent with a short-range entangled Anderson-localized state. 
Note that this argument is closely analogous to the Chalker-Coddington-type argument for delocalization at the 2D quantum Hall plateau transition (or on the surface of the 3D TR-invariant TI). 
This is not an accident, but expresses the fact that the magnetic Weyl semimetal may be viewed as a 3D generalization of the 2D plateau transition, as already implied by the fact that 
the 2D branch surfaces of the phase $\theta$ are decorated by 2D Chern insulators. For a more detailed discussion of this see Ref.~\cite{Yi25}. 

Now let us see if this result generalizes to the case of TR-invariant Weyl semimetal, with topological response described by Eq.~\eqref{eq:13}.  
Following the same logic as above, we make a replacement
\beq
\label{eq:24}
2 \delta Q a e^z \ra d \theta, 
\eeq
where $\theta$ is the phase of the translational symmetry breaking mass term for a pair of Weyl nodes. Note that since the two pairs are related to each 
other by TR symmetry, the two masses must also be related by TR and described by a single complex translation-breaking field. 
This gives the action
\beq
\label{eq:25}
S = i \frac{2 Q a}{4 \pi^2} \int d \theta \wedge d e^z \wedge A. 
\eeq
Taking $dA = 0$, one obtains for the action of a vortex line
\beq
\label{eq:26}
S = i \frac{2 Q a}{2 \pi} \int e^z \wedge A, 
\eeq
which corresponds to a 1D metal with the Fermi points at $k_z = \pm Q$~\cite{Wang21}. 

Unlike the case of the magnetic Weyl semimetal above, where the vortex bound state was a chiral metal, here one seems to obtain an ordinary 1D metal, which is 
not immune to localization. 
However, this conclusion is too hasty: this 1D metal is actually not ordinary, but helical, since the states at the two Fermi points are an exact Kramers doublet. 
This prohibits backscattering, and therefore localization, as long as the TR symmetry is strictly preserved (i.e. disorder is nonmagnetic). 

As in the magnetic Weyl semimetal case, statistical restoration of the translational symmetry requires a system-spanning vortex network. 
One might worry here that backscattering between the right and left-moving modes becomes possible once distinct vortex lines approach closely, so tunneling between them is not negligible
(right- and left-movers localized on different vortex lines are not Kramers partners). 
However, we may use exactly the same coarse-graining argument as in the magnetic Weyl semimetal case, replacing chiral modes by helical modes, to argue that localization 
is impossible if translational symmetry is restored statistically. 
Thus, as in the magnetic Weyl semimetal case above, translational symmetry pins the helical network in the delocalized phase. 
Note that this argument for the lack of localization in TR-invariant Weyl semimetals is closely analogous to the argument for delocalization of the surface states of 3D TR-invariant 
weak topological insulators~\cite{Fu12}. 

Now let us proceed to the type-I Dirac semimetal case, where topological response is described by Eq.~\eqref{eq:14}. 
Considering a rotation symmetry breaking mass term with the phase $\theta$ (see Ref.~\cite{Gioia21} for a microscopic model), we make a replacement
\beq
\label{eq:27} 
\frac{2 \pi q}{n} c_n \ra d \theta,
\eeq
which gives
\beq
\label{eq:28}
S = - i \frac{\lambda}{2 \pi} \int d \theta \wedge e^z \wedge d A. 
\eeq
Importantly, rotational symmetry acts only discretely on the phase $\theta$, unlike the effectively continuous translation-induced phase rotations considered above.
The topological defects in this case are domain walls between regions with different values of the phase and the details of the calculations depend significantly on the 
specific case. We will thus explicitly consider the Dirac semimetal realization from Ref.~\cite{Gioia21}, with Dirac points protected by the $C_4$ symmetry. 
This type-I Dirac semimetal may be viewed as an intermediate phase between a weak 3D 
TR-invariant TI and an ordinary insulator, where a transition point has been broadened into an intermediate phase by an extra rotational symmetry. 
While we do not have a rigorous proof, we believe that the conclusions are generic for all type-I Dirac semimetals protected by rotational symmetry. 

In this case we have $n = 4$ and $q = 2$, i.e. the rotation symmetry breaking mass changes sign under the $C_4$ rotation, which means that we may assume an ensemble in 
which $\theta$ only takes two values, $0$ and $\pi$. 
Taking $d e^z = 0$, we obtain the domain wall action
\beq
\label{eq:29}
S = - i \frac{\lambda}{2} \int e^z \wedge d A. 
\eeq
While Eq.~\eqref{eq:29} has the same form as the polarization response Eq.~\eqref{eq:17}, its symmetry and anomaly interpretations are very different. 
Indeed, unlike Eq.~\eqref{eq:17}, the response of Eq.~\eqref{eq:29} is not anomalous on the domain wall. In the 2D Dirac semimetal of 
Eq.~\eqref{eq:17}, mirror symmetry reverses the polarization and makes a continuously tunable polarization incompatible with a mirror-symmetric short-range-entangled state. 
On the present $C_4$-breaking domain wall, the remaining symmetries do not quantize the polarization. Eq.~\eqref{eq:29} is therefore compatible with a stand-alone short-range-entangled 2D system. The domain walls may consequently proliferate without requiring delocalized electronic states, and the type-I Dirac semimetal is therefore not protected from Anderson localization.
Note that this is not in conflict with the fact that the clean Dirac semimetal may intervene between topologically distinct insulating phases: the distinction between the latter requires delocalized states somewhere along the interpolation, but does not require a finite delocalized phase.
\subsection{Unquantized parity anomaly}
\label{sec:3.2}
Let us now consider the final two cases with the unquantized parity anomaly. 
Let us start with the case of the 2D Dirac semimetal with a pair of nodes, protected by mirror symmetry. 
Its unquantized anomaly is given by Eq.~\eqref{eq:17}. 
\begin{table*}[t]
\centering
\small
\setlength{\tabcolsep}{3pt}
\renewcommand{\arraystretch}{1.35}
\newcommand{\tcell}[2]{
  \parbox[c][0.55in][c]{#1}{
    \centering
    \linespread{1}\selectfont
    #2
  }
}
\begin{tabular}{|c|c|c|c|c|c|}
\hline
\tcell{0.15\textwidth}{\textbf{System}} &
\tcell{0.17\textwidth}{\textbf{Unquantized anomaly}} &
\tcell{0.17\textwidth}{\textbf{Mass term}} &
\tcell{0.13\textwidth}{\textbf{Defect}} &
\tcell{0.15\textwidth}{\textbf{Defect response}} &
\tcell{0.13\textwidth}{\textbf{Protection from localization}} \\
\hline \hline

\tcell{0.15\textwidth}{Magnetic Weyl semimetal} &
\tcell{0.17\textwidth}{$e^z \wedge A \wedge dA$} &
\tcell{0.17\textwidth}{Complex translation-breaking mass} &
\tcell{0.13\textwidth}{Vortex line} &
\tcell{0.15\textwidth}{Chiral 1D metal} &
\tcell{0.13\textwidth}{Yes} \\
\hline

\tcell{0.15\textwidth}{TR-invariant Weyl semimetal} &
\tcell{0.17\textwidth}{$e^z \wedge d e^z \wedge A$} &
\tcell{0.17\textwidth}{Complex translation-breaking mass} &
\tcell{0.13\textwidth}{Vortex line} &
\tcell{0.15\textwidth}{Helical 1D metal} &
\tcell{0.13\textwidth}{Yes} \\
\hline

\tcell{0.15\textwidth}{Type-I Dirac semimetal} &
\tcell{0.17\textwidth}{$e^z \wedge c_n \wedge dA$} &
\tcell{0.17\textwidth}{Complex rotation-breaking mass} &
\tcell{0.13\textwidth}{2D domain wall} &
\tcell{0.15\textwidth}{Short-range-entangled insulator} &
\tcell{0.13\textwidth}{No} \\
\hline

\tcell{0.15\textwidth}{2D Dirac semimetal} &
\tcell{0.17\textwidth}{$e^x \wedge dA$} &
\tcell{0.17\textwidth}{Complex translation-breaking mass or scalar mirror-breaking mass}&
\tcell{0.13\textwidth}{Point vortex or 1D domain wall} &
\tcell{0.15\textwidth}{Half-charge or 1D metal} &
\tcell{0.13\textwidth}{No} \\
\hline

\tcell{0.15\textwidth}{3D nodal line semimetal} &
\tcell{0.17\textwidth}{$e^x \wedge e^y \wedge dA$} &
\tcell{0.17\textwidth}{Scalar mirror-breaking mass} &
\tcell{0.13\textwidth}{2D domain wall} &
\tcell{0.15\textwidth}{2D metal} &
\tcell{0.13\textwidth}{No} \\
\hline
\end{tabular}

\caption{Unquantized anomalies, associated mass terms, defects, defect responses, and localization protection in topological semimetals.}
\label{tab:1}
\end{table*}
Such a Dirac semimetal may be gapped by a complex translation-breaking mass term that mixes the two nodes and carries momentum $2 Q$. 
Taking $\theta$ to be the phase of the mass term, we make a replacement 
\beq
\label{eq:30}
2 Q a e^x \ra d \theta. 
\eeq
This gives the action for the vortex 
\beq
\label{eq:31}
S = \pm \frac{i}{2} \int A.
\eeq
This describes a fractional charge $\pm 1/2$ in the vortex core. 
Restoring the broken translational symmetry at long length scales requires creating such vortices with half-quantized charges in the core. 
However, since the vortices (which are points rather than lines in this case) may be arbitrarily far apart in a macroscopic sample, there is no analog of the percolation of gapless modes,
and this generally corresponds to a localized state. 
This means that the unquantized anomaly response Eq.~\eqref{eq:17} is compatible with an Anderson insulator, in contrast to the 3D Weyl semimetal case. 

Alternatively, it is also instructive to consider mirror symmetry breaking disorder instead. 
The 2D Dirac semimetal is protected by mirror symmetry and may be gapped by a scalar symmetry-breaking mass. 
We may view the sign in front of Eq.~\eqref{eq:17} as the sign of this scalar mass term. 
Taking $d e^x = 0$, this leads to the following action for a 1D domain wall between two gapped states, corresponding to the opposite mass signs
\beq
\label{eq:31.5}
S = \pm i \lambda \int e^x \wedge A. 
\eeq
This corresponds to an ordinary 1D electron or hole-like metal with fractional charge per unit cell $\lambda$ (not helical as the TR symmetry is explicitly broken). 
Such a 1D metal is not protected from localization, which again implies that the 2D Dirac semimetal is also not protected. 

Finally, let us consider the case of the 3D nodal line semimetal, protected by mirror symmetry. 
The unquantized anomaly is given by Eq.~\eqref{eq:18}. 
Just as in the case of the 2D mirror-protected Dirac semimetal, the 3D nodal line semimetal may be fully gapped by a scalar mirror-symmetry breaking mass, see Ref.~\cite{Wang21} for a concrete microscopic model. 
As before, we may view the sign in front of Eq.~\eqref{eq:18} as the sign of this scalar mass term. 
Taking $d e^x = d e^y = 0$, this leads to the following action for a planar 2D domain wall between two gapped states, corresponding to the opposite mass signs
\beq
\label{eq:32}
S = \pm i \lambda \int e^x \wedge e^y \wedge A. 
\eeq
This corresponds to a 2D electron or hole-like metal with fractional charge per unit cell $\lambda$. 
Again, such a 2D metal in the unitary symmetry class is not protected from localization, which implies that the nodal line semimetal is also not protected and the unquantized anomaly Eq.~\eqref{eq:18} is again compatible with a short-range entangled Anderson insulator. 
Thus fractional polarization in a mirror symmetric system only requires a long-range entangled state in the absence of disorder. 
An Anderson insulator with statistical mirror symmetry may have any polarization and the decorated domain wall approach provides an explicit construction of such an insulator. 
Our results are summarized in Table~\ref{tab:1}. 

\section{Discussion and conclusions}
\label{sec:4}
We have addressed the question of whether the nontrivial topology of a clean TM survives in the presence of disorder. The main point is that, in a disordered system, this question cannot be formulated in terms of topology-enforced spectral gaplessness. Disorder generically fills spectral gaps with localized states, so the distinction between a gapless and a gapped spectrum is no longer applicable. The more meaningful question is instead whether the state can become short-range entangled, i.e. Anderson localized, while preserving the corresponding topological response in the average sense.

The answer we have found is nontrivial. Among the standard examples of TM, only 3D Weyl semimetals (both magnetic and TR-invariant) have topological responses that are incompatible with a short-range entangled Anderson insulator. When disorder locally breaks translation symmetry and generates a random mass, gapping the Weyl nodes, restoring translation symmetry on long length scales requires proliferation of vortices in the phase of this mass. The dimensional anomaly descent of the bulk response implies that these vortices bind 1D chiral or helical modes. 
Since the vortex defects carry anomalous chiral or helical modes that cannot themselves be localized, the average-symmetry argument of Section~\ref{sec:3} excludes a statistically symmetric state with a finite localization length.
Thus the unquantized anomaly of 3D Weyl semimetals enforces long-range entanglement.

We have demonstrated that this conclusion does not extend to other TM, in particular to type-I 3D Dirac semimetals as well as mirror-protected 2D Dirac and 3D nodal line semimetals. 
In these cases the descended defect responses are compatible with short-range-entangled states, or with lower-dimensional metals that are themselves not protected from localization. Thus the proliferation of the symmetry-restoring defects does not force delocalized electronic states to survive at long distances.

The distinction may be summarized as follows. An unquantized anomaly by itself does not guarantee delocalization. 
What matters is whether dimensional descent produces a defect response that is itself incompatible with a short-range-entangled localized state.
This condition is satisfied for 3D Weyl semimetals, but not for other types of topological semimetals, considered in this paper. 

This also clarifies the relation between TM and average SPT. In average SPT phases, the decorated-domain-wall construction shows that the distinction between the SPT and a trivial insulator may survive disorder even when the protecting symmetry is violated in each individual disorder realization, provided the symmetry is restored statistically and provided the corresponding 
domain wall states are higher than zero-dimensional~\cite{Ma22}. For TM the analogous criterion is stronger. It is not sufficient for the domain walls or defects to carry nontrivial quantum numbers and to percolate. They must carry modes capable of delocalized transport that cannot itself be removed by disorder. 

Our discussion has explicitly not included type-II Dirac semimetals, where Dirac points, pinned to TR-invariant momenta in the BZ, are protected by nonsymmorphic 
symmetries~\cite{Kane12,Steinberg14,Gioia25}.
Since nonsymmorphic symmetries involve both point-group operations and translations, and since such semimetals are not described by the unquantized anomalies, 
considered here, their localization properties cannot be inferred from the present results and will be left for future work.

To conclude, let us highlight why lack of localization in Weyl semimetals is unusual from a slightly different angle. 
In an ordinary 3D metal, the diagonal conductivity is given by
\beq
\label{eq:33}
\sigma_{xx} \sim \frac{e^2}{h} k_F (k_F \ell), 
\eeq
where $k_F$ is the Fermi wavevector and $\ell$ is the mean free path. 
As the Fermi energy is reduced towards the bottom of the band, one enters the ``bad metal" regime when $k_F \ell \lesssim 1$, at which point, at least in a weakly interacting system, 
an Anderson localization transition is generally expected. 

In a Weyl semimetal $k_F \ra 0$ and~\cite{Altland16,Burkov26-1}
\beq
\label{eq:34}
\sigma_{xx} \sim \frac{e^2}{h} \frac{1}{\ell}, 
\eeq
which gives the conductance at the scale of the mean free path
\beq
\label{eq:35}
g_{xx}(\ell) = \sigma_{xx} \ell \sim \frac{e^2}{h}. 
\eeq
This means that there is no large bare conductance parameter, protecting the metallic state, and one could generally expect localization
(whether a particular sample is above or below the localization threshold will depend on microscopic details). However, localization is in fact avoided due to nontrivial topology. 
This is very similar to how localization is avoided at the integer quantum Hall plateau transition in 2D and there does exist a close analogy between a
3D magnetic Weyl semimetal and a 2D quantum Hall plateau transition~\cite{Yi25}, as already mentioned above. 
This lack of localization in a 3D magnetic Weyl semimetal is consistent with the observations in Ref.~\cite{Belopolski25}. 
The MBE-grown Cr, In and Sb-doped Bi$_2$Te$_3$ samples of Ref.~\cite{Belopolski25} are strongly disordered,
yet exhibit robust metallic transport, even though the diagonal resistivity is very high for a metal. 

\begin{acknowledgments}
We thank Chong Wang for useful discussions. 
Financial support was provided by the Natural Sciences and Engineering Research Council (NSERC) of Canada.
Research at Perimeter Institute is supported in part by the Government of Canada through the Department of Innovation, Science and Economic Development and by the Province of Ontario through the Ministry of Economic Development, Job Creation and Trade.
\end{acknowledgments}
\bibliography{references}
\end{document}